\documentclass[amsmath,amssymb, 12pt, nofootinbib]{revtex4}
\usepackage{graphicx}
\usepackage{dcolumn}
\usepackage{bm}
\usepackage{graphicx}
\usepackage{epsfig}

\def\bea{\begin{eqnarray}}
\def\eea{\end{eqnarray}}

\begin{document}

\title{Holographic dark energy interacting with dark matter in a Closed Universe}
\author{Norman Cruz$^{1}$, Samuel Lepe$^{2}$, Francisco Pe\~na$^{3}$ and Joel Saavedra $^{2}$}
\address{$^{1}${\small Departamento de
F\'\i sica, Facultad de Ciencia, Universidad de Santiago, Casilla
307, Santiago, Chile.}}
\address{ $^{2}${\small Instituto de F\'{\i}sica, Pontificia
Universidad Cat\'olica de Valpara\'{\i}so, Casilla 4950,
Valpara\'{\i}so,} }
\address{ $^{3}${\small Departamento de
Ciencias F\'\i sicas, Facultad de Ingenier\'\i a, Ciencias y
Administraci\'on, Universidad de La Frontera, Avda. Francisco
Salazar 01145, Casilla 54-D Temuco, Chile.} }

\date{\today}
\begin{abstract}
A cosmological model of an holographic dark energy interacting
with dark matter throughout a decaying term of the form $Q=3\left(
\lambda _{1}\rho _{DE}+\lambda _{2}\rho _{m}\right) H$ is
investigated. General constraint on the parameters of the model
are found when accelerated expansion  is imposed and we found a
phantom scenarios, without any reference to a specific equation of
state for the dark energy. The behavior of equation of stated for
dark energy is also discussed.

\end{abstract}
\maketitle

\section{Introduction}

In the recent years a number of observational facts from high
redshift surveys of type Ia Supernovae, WMAP, CMB, etc, led us to
believe that our universe is passing through an accelerating phase
of expansion~\cite{Riess}. It is generally accepted that our
universe might have also emerged from an accelerating phase in the
past. Thus there might have two phases of acceleration of the
universe, early inflation and late acceleration followed by a
decelerating phase. In order to describe the present accelerating
phase of the universe, it may be useful to consider dark energy in
the theory. The observational data of the universe indicates that
dark energy content of the universe is about $76\%$ of the total
energy budget of the universe. To accommodate such a huge energy
various kinds of exotic matters are considered to identify
possible candidate for the dark energy. Recently, holographic
principle~\cite{Fischler},~\cite{Bousso} is incorporated in
cosmology~\cite{Hsu},~\cite{Li},~\cite{Zhang}, ~\cite{Huang} to
constraint the dark energy content of the universe following the
work of Cohen et al.~\cite{Cohen:1998zx}. The holographic
principle, in simple words establish that all degrees of freedom
of a region of space in are the same as that of a system of binary
degrees of freedom distributed on the boundary of the region
\cite{Susskind:1994vu}. This point of view, represents an approach
from a consistent theory of quantum gravity (unfortunately not yet
found) in order to clarify the nature of dark energy.  The
Holographic principle says that the number of degrees of freedom
of a physical system should scale with its bounding area rather
than with its volume. Along these lines the literature have been
focused in explain the size of the dark energy density on the
basis of holographic ideas, derived from the suggestions that in
quantum field theory a short distance  a cut-off is related to a
long distance cut-off due to the limit set by the formation of a
black hole \cite{Cohen:1998zx}. Interacting dark energy and dark
matter approach has been extensively discussed in the literature,
see Ref. \cite{He:2008tn} and references therein. One of the most
used description for the interaction between dark energy and dark
matter is describe through an interaction factor given by
$Q\sim\left(\lambda_1\,\rho_{DE}+\lambda_2\,\rho_{DM}\right)H $,
whose origin can be modelled from a phenomenological point of view
\cite{Wang:2006qw, Wang:2005ph, Wang:2005pk, He:2008tn} and front
observational data \cite{Quartin:2008px}. In this article we are
going to consider dark energy from an holographic origin.

The plan of the paper is as follows: In Sec. II we described the
interacting model, accelerated scenarios and phantom regimen. In
Sec. III we discuss the behavior of the equation of state.
Finally, we conclude in Sec. IV.

\section{Dark Energy Decaying to Dark Matter}
In the following we modelled the universe assuming that is filled
with dark matter (including both dark and barionic matter), with a
density $\rho_{m}$, and a dark energy component, $\rho_{DE}$,
which obey the holographic principle. We are going to assume that
the dark matter component is interacting with the dark energy
component through a source (loss) term $Q$ that enters the energy
balance. The Friedmann's field equation is given by
\begin{eqnarray}\label{friedmann00}
3H^{2}+\frac{3k}{a_{0}^{2}}\chi ^{-2} &=&\rho _{DE}+\rho _{m},
\end{eqnarray}
The continuity equations for both fluids take the form
\begin{eqnarray}\label{continuity1}
\dot{\rho}_{DE}+3H\left( \rho _{DE}+p_{DE}\right) &=&-Q,
\end{eqnarray}
\begin{eqnarray}\label{continuity2}
\dot{\rho}_{m}+3H\rho _{m} &=&Q,
\end{eqnarray}
where $\chi =a/a_{0}$ \footnote{$a_{0}$ is the present day value
of the scale factor.} and 8$\pi G=1$. Dark matter obey the
equation of state corresponding to dust $p_{m}=0$. In the
following we shall use the interaction model given by
\begin{eqnarray}\label{Qinteraction}
Q=3\left( \lambda _{1}\rho _{DE}+\lambda _{2}\rho _{m}\right) H.
\end{eqnarray}
where $\lambda _{1}$ and $\lambda _{2}$ are positive constants.
This model are basic in a phenomenologically way from the
interaction between dark energy and dark matter, and were
extensively considered in the literature, see Refs.
\cite{Wang:2006qw, Wang:2005ph, Wang:2005pk, He:2008tn} and
references therein. Substituting Eq.~(\ref{Qinteraction}) in
Eq.~(\ref{continuity2}), we obtain the corresponding coupled
equation
\begin{eqnarray}\label{continuity2a}
\dot{\rho}_{m}+3\left( 1-\lambda _{2}\right) H\rho _{m}=3\lambda
_{1}\rho _{DE}H.
\end{eqnarray}
We now assume that the holographic dark energy takes the form
\begin{eqnarray}\label{holomodel}
\rho _{DE}=3c^{2}H^{2},
\end{eqnarray}
different cases with $c^2\lessgtr 1$ were well described from
theoretical point of view in Ref. \cite{Huang:2004ai} and from the
observational view in Ref. \cite{Huang:2004wt}. Substituting the
expression Eq.~(\ref{holomodel}) in Eq.~(\ref{continuity2a}), and
using Eq.~(\ref{friedmann00}), we obtain the following solution
for $\rho _{m}$
\begin{eqnarray}\label{rho2}
\rho _{m}\left( \chi \right) =\left[ C_{1}\chi ^{\Delta
}+k\frac{9c^{2}\lambda _{1}}{\left( 1-c^{2}\right) a_{0}^{2}\Delta
}\right] \chi ^{-2},
\end{eqnarray}
where $C_{1}$ is a positive constant of integration, and $\Delta$
is a constant defined by
\begin{eqnarray}\label{DELTA}
\Delta \equiv 3\left(\lambda _{2}+\frac{c^{2}}{1-c^{2}}\lambda
_{1}\right)-1 .
\end{eqnarray}
The expression given by Eq.~(\ref{rho2}) reproduces in the limit
with no interaction the usual behavior for dust, i.e., $\rho
_{m}\sim \chi^{-3}$. Nevertheless, it is straightforward to see
that Eq.~(\ref{continuity2a}) is equivalent to the continuity
equation of a fluid with an equation of state
$p_{eff}=-(\lambda_{1}/r + \lambda_{2})\rho _{m}$, where $r$ is
the coincidence parameter $r=\rho_{m}/\rho_{DE}$ . In other words,
due to the interaction with dark energy the dark matter behaves
like a fluid with negative pressure.

In order to obtain an expression for the Hubble parameter, we
introduce in Eq.~(\ref{friedmann00}) the holographic dark energy
for $\rho _{DE}$, and from Eq.~(\ref{rho2}), yields
\begin{eqnarray}\label{H2}
H^{2}\left( \chi \right) &=&\frac{1}{3\left( 1-c^{2}\right)
}\left[ C_{1}\chi ^{\Delta }-k\delta \right] \chi ^{-2}.
\end{eqnarray}
As we shall see bellow, a closed universe ($k=1$) implies a non
constant coincidence parameter with the cosmic time. Since
$H^{2}>0$ we choose $\delta <0$. Taking the derivative with
respect to the cosmic time of the above equation yields
\begin{eqnarray}\label{Hpoint}
\dot{H}\left( \chi \right) &=&\frac{1}{6\left( 1-c^{2}\right)
}\left[ C_{1}\left( \Delta -2\right) \chi ^{\Delta }+2k\delta
\right] \chi ^{-2},
\end{eqnarray}
where $\delta$ is a constant defined by
\begin{eqnarray}\label{delta}
\delta \equiv \frac{3}{a_{0}^{2}}\left( 1-\frac{3c^{2}}{\left(
1-c^{2}\right) \Delta }\lambda _{1}\right).
\end{eqnarray}
From Eqs.~(\ref{H2}) and (\ref{Hpoint}) and since $H^{2}\left(
\chi \right) +\dot{H}\left( \chi \right) =\frac{\ddot{\chi}}{ \chi
}$, we obtain the expression for the acceleration
\begin{eqnarray}\label{acceleration}
\ddot{\chi}=\frac{C_{1}}{6\left( 1-c^{2}\right) }\Delta \chi
^{\Delta -1}.
\end{eqnarray}
The decceleration parameter $q$ is given by
\begin{eqnarray}\label{decceleration}
q=-\frac{\ddot{a}}{aH^{2}}=- \left( 1+ \frac{\dot{H}}{H^{2}}
\right),
\end{eqnarray}
and from Eqs.~(\ref{Hpoint}) and ~(\ref{H2}) we obtain that
\begin{eqnarray}\label{qqq}
q(\chi)=- \left( 1+ \frac{1}{2}\frac{\left[ C_{1}\left( \Delta
-2\right) \chi ^{\Delta }+2k\delta \right]}{C_{1}\chi ^{\Delta
}-k\delta } \right),
\end{eqnarray}

\subsection{An accelerated universe}

We investigate first the conditions to have an accelerating
universe, imposing that $\ddot{\chi}>0$. Eq.~(\ref{acceleration})
tell us that  $\Delta$ must be positive. Now we  are going to
analyze the quantity $\Delta$ according to the constraints
$0<\Delta<2$ and $\Delta \geq2$  for $\delta<0$.  It is
straightforward to obtain the following constraints for
$\lambda_1$ and $\lambda_2$. In the former case
\begin{eqnarray}\label{labmda}
\frac{1-c^2}{c^2}&&\left(\frac{1}{3}-\lambda_2\right)<\lambda_1<\frac{1-c^2}{c^2}\left(1-\lambda_2\right),
\nonumber\\&& 0<\lambda_2<\frac{1}{3},
\end{eqnarray}
and when $\Delta \geq2$,
\begin{eqnarray}\label{labmda11}
\lambda_1&&>\frac{2}{3}\,\left( \frac{1-c^2}{c^2} \right),
\nonumber\\&& 0<\lambda_2<\frac{1}{3}.
\end{eqnarray}
In these universes with accelerated expansion the dark matter
component which, as we have seen before (see Eq.(\ref{rho2})), has
a negative pressure, behaves like the sum of two fluids both with
decreasing energy density when the scale factor grows (if
$0<\Delta<2$). On the other hand, if $\Delta>2$ both dark
components diverge when $\chi$ grows.

In our model it is easy to check that the coincidence parameter
is a decreasing function of the scale between two extreme values.
The expression for $r$, evaluated from Eqs.~(\ref{holomodel}) and
~(\ref{rho2}), yields
\begin{eqnarray}\label{rparameter}
r(z)=\frac{1-c^2}{c^2}\left[1+\frac{3k}{a_{0}^2}\frac{1}
{C_{1}(1+z)^{-\Delta}-k\delta}\right],
\end{eqnarray}
where $1+z={\chi}^{-1}$. Deriving the parameter $r$ with respect
to the variable  $\chi$, we obtain the following expression in
terms of the acceleration
\begin{eqnarray}\label{rparameter1}
\frac{dr(\chi)}{d\chi}= -\frac{18k(1-c^2)^{2}}{c^2 a_{0}^2}
\frac{\ddot{\chi}}{(C_{1}\chi^{\Delta}-k\delta)^{2}}.
\end{eqnarray}
Note that if we have imposed an accelerated expansion
($\ddot{\chi}>0$) a decreasing $r$ is obtained only for $k>0$. So
in our approach open universe leads to not desirable physical
results. Evaluating the limits of $r$ for $r(z\rightarrow \infty)$
($\chi\rightarrow 0$) and $r(z\rightarrow -1)$ ($\chi\rightarrow
\infty$) we obtain
\begin{eqnarray}\label{rlimits}
r(z\rightarrow \infty)\rightarrow
\frac{\lambda_{1}}{1/3-\lambda_{2}}; \,\,\,\,\,\,r(z\rightarrow
-1)\rightarrow \frac{1-c^2}{c^2}.
\end{eqnarray}
For flat universes $r(z)$ is always constant. In the case of a
closed universe $r$ goes to a constant in the future cosmic
evolution.

In order to analyze the behavior of the interaction term $Q$ we
rewrite Eq.~(\ref{Qinteraction}) in the form
\begin{eqnarray}\label{Qinteraction1}
Q=3\left( \lambda _{1}+\lambda _{2}r(z)\right)\rho_{DE}H(z),
\end{eqnarray}
and explicitly for our holographic model the interaction takes the
form
\begin{eqnarray}\label{Qinteraction2}
Q(z)=9c^2\left(\lambda_1+\lambda_2\,r(z)\right)H^3(z).
\end{eqnarray}
If the only condition on $\Delta$ is to be positive (with $\delta
<0$) we obtain a reasonable physical result since the behavior of
$Q$ as the cosmic time evolves is a decreasing function. In the
limit $z\rightarrow-1$, we obtain $Q(z\rightarrow-1)\rightarrow
0$, since $H(z\rightarrow-1)\rightarrow 0$. Meanwhile, the dark
energy density (see Eq.~(\ref{rho2})) is also a decreasing
function of the cosmic time, which justify a decreasing
interaction between the two fluids considered.  This scenario
occurs if $0<\Delta \leq2$. If $\Delta >2$ we have divergences in
the interaction term and in the dark components.

\subsection{The phantom case}

It is interesting to note that phantom scenarios are allowed by
Eq.~(\ref{acceleration}) since admit solutions of the type $\chi
\left( t\right) =\left( t_{s}-t\right) ^{-\beta}$ with $\beta>0$.
In the phantom scenarios where $\Delta>2$, we have a divergence
for the scale factor $\chi$ and in the Hubble parameter, since
$H(z\rightarrow-1)\rightarrow (1+z) ^{1-\Delta/2}$. The
singularity of the scale factor occurs at $t=t_{s}$, which can be
evaluated integrating twice Eq.(\ref{acceleration}). We obtain
that the cosmic time is given by
\begin{eqnarray}\label{tsingularity}
t+A_{2}=\int \frac{d\chi}{\sqrt{\frac{C_{1}\chi
^{\Delta}}{3(1-c^{2})}+ A_{1}}}.
\end{eqnarray}
Choosing for simplicity, $A_{1}=0$ and fixing $A_{2}$ in order to
have $\chi(t_{0})=1$, we obtain
\begin{eqnarray}\label{tsingularity}
\chi^{\Delta/2 -1}=
\frac{1}{\sqrt{\frac{C_{1}}{3(1-c^{2})}}(\frac{\Delta}{2}
-1)}\left(t_{s}-t\right)^{-1}.
\end{eqnarray}
The expression for $t_{s}$ is then given by
\begin{eqnarray}\label{tsingularity}
t_{s}=t_{0}+\frac{1}{\sqrt{\frac{C_{1}}{3(1-c^{2})}}(\frac{\Delta}{2}
-1)}.
\end{eqnarray}
At this future time we obtain the usual singularities that
characterize a big rip solution. We note that for the current time
from Eq. (\ref{qqq}) it is straightforward to obtain
\begin{equation}
\label{qqqq} q(0)>-\frac{1}{2}\Delta,
\end{equation}
in accord to the current observational data \cite{Seljak:2006bg}.
\section{The behavior of the equation of state}

In this section we investigate the equation of state of the dark
energy, which has been only restricted by the holographic
criteria and by its interaction with dark matter. In doing so, we
reduce the two fluid components to an equivalent one fluid with
an equation of state $p=\omega\rho$ and pressure $p$. We make
this by taking the equations of state
$p_{DE}=\omega_{DE}\rho_{DE}$ and $p_{m}=0$, and replacing the
Eq.~(\ref{continuity1}) and Eq.~(\ref{continuity2}). The sum of
these equations yields
\begin{eqnarray}
\dot{\rho}+3H(1+\omega)\rho=0
\end{eqnarray}
where $\rho=\rho_{DE}+\rho_{m}$ and
$\omega\rho=\omega_{DE}\rho_{DE}$.

The Friedmann equation
becomes
\begin{eqnarray}\label{sourcedFridman}
3H^{2} =\rho-\frac{3k}{{a_{0}}^2}\chi^{-2}.
\end{eqnarray}
It is straightforward to obtain the following expression for
$\omega$
\begin{equation}\label{eqforomega}
1+\omega=-\frac{2}{3}\left[\frac{\dot{H}-
\frac{k}{{a_{0}}^{2}}{\chi}^{-2}}{H^2+\frac{k}{{a_{0}}^2}{\chi^{-2}}}\right].
\end{equation}
Using the expression for $\rho_{DE}$ and $Q$ yields
\begin{eqnarray}\label{stateparameter2}
1+\omega=-\frac{1}{3}\left[\frac{C_{1}(\Delta
-2){\chi}^{\Delta}+2k\beta}{C_{1}{\chi}^{\Delta}-k\beta}\right],
\end{eqnarray}
with
$\beta=\delta-\frac{3(1-c^2)}{{a_{0}}^2}=\frac{3}{{a_{0}}^{2}}\left[1-\frac{3c^2
\lambda_{1}}{\Delta}\right]$. From the definition of $\omega$ we
obtain
\begin{eqnarray}\label{omegarDE}
\omega (1+r)=\omega_{DE}\;\;.
\end{eqnarray}
Using in the above equation the expression for $r$ given by
Eq.(\ref{rparameter}) we obtain the equation of state for the
dark energy fluid
\begin{eqnarray}\label{omegaDE}
\omega_{DE}=-\frac{1}{3c^2}\left[1+\Delta\frac{C_{1}{\chi}^{\Delta}}{C_{1}{\chi}^{\Delta}-k\beta}\right]
\end{eqnarray}
Without an explicit identification of a particular model for the
dark energy, it is of interest to evaluate the limit
$\omega(z\longrightarrow -1)$ and evaluate
$\omega(z\longrightarrow \infty)$, in order to know the future and
early behavior, respectively,  of the effective fluid of the
universe. For the future we obtain
\begin{eqnarray}\label{limw1}
\omega(z\longrightarrow -1)\longrightarrow -1-\frac{1}{3}(\Delta
-2),
\end{eqnarray}
where, as we mentioned before, $\Delta>2$ implies phantom
behavior. The cosmological constant is obtained for $\Delta=2$ and
for early times yields
\begin{eqnarray}\label{limw3}
\omega(z\longrightarrow \infty)\longrightarrow -\frac{1}{3}.
\end{eqnarray}
From Eq.(\ref{omegaDE}) we obtain the expression for
$\omega_{DE}(z)$, which is given by
\begin{eqnarray}\label{omega1z}
\omega_{DE}(z)=-\frac{1}{3c^2}\left[1+\frac{\Delta}{1-k\beta
{C_{1}}^{-1}(1+z)}\right].
\end{eqnarray}
As the cosmic time evolves from now, the equation of state of the
dark energy behaves like
\begin{eqnarray}\label{limw11}
\omega_{DE}(z\longrightarrow -1)\longrightarrow
-\frac{1+\Delta}{3c^2},
\end{eqnarray}
and at early times the dark energy becomes
\begin{eqnarray}\label{limw13} \omega_{DE}(z\longrightarrow
\infty)\longrightarrow -\frac{1}{3c^2}.
\end{eqnarray}

Those results show that the dark energy evolves from a near
behaviors like as stringy gas to a phantom behavior.

\section{Discussion}

A cosmological model of an holographic dark energy interacting
with dark matter throughout a decaying term of the form $Q=3\left(
\lambda _{1}\rho _{DE}+\lambda _{2}\rho _{m}\right) H$ in a closed
universe is investigated. General constraint on the parameters of
the model are found when accelerated expansion is imposed. A
dynamical coincidence parameter was found with dynamical decaying
in the late times, that it is consistent with the constraint
$c^2<1$. The dynamical parameters $H$, $Q$, $\rho_{DE}$ and
$\rho_{DM}$ have the adequate limits at early and late times. We
found that a phantom solutions is allowed, because
Eq.~(\ref{acceleration}) admit solutions of the type $\chi \left(
t\right) =\left( t_{s}-t\right) ^{-\beta}$ with $\beta>0$. Here we
obtain a deceleration parameter given by $q(0)>-\frac{1}{2}\Delta$
with $\Delta>2$ and this constraint on $\Delta$ is consistent with
the observational data, without any reference to a specific
equation of state for the dark energy. At this point, we
investigated the behavior of equation of state of the dark energy,
which was only restricted by the holographic criteria and by its
interaction with dark matter. We found, that the early limit is
described by $\omega_{DE}(z\longrightarrow -1)\longrightarrow
-\frac{1+\Delta}{3c^2}$ and the late times is given by
$\omega_{DE}(z\longrightarrow \infty)\longrightarrow
-\frac{1}{3c^2}$. Those results show that the dark energy evolves
from a near behaviors like as stringy gas to a phantom behavior.

\section{Acknowledgements} NC, SL and JS acknowledge the hospitality of
the Physics Department of Universidad de La Frontera where part of
this work was done. SL acknowledges the hospitality of the Physics
Department of Universidad de Santiago de Chile. FP acknowledges
the hospitality of the Physics Institute of Pontificia Universidad
Cat\'olica de Valpara\'\i so. We acknowledge the partial support
to this research by CONICYT through grant N$^0$ 11060515 (JS) and
by Direcci\'on de Estudios Avanzados PUCV. It also was supported
from DICYT 040831 CM, Universidad de Santiago de Chile (NC),
DIUFRO DI08-0041, of Direcci\'on de Investigaci\'on y Desarrollo,
Universidad de La Frontera (FP) and DI-PUCV, Grants 123.701/08
(SL) and 123.789 (JS), Pontificia Universidad Cat\'olica de
Valpara\'\i so.

\end{document}